# Novel metal-insulator-transition at the SrTiO$_3$/SmTiO$_3$ interface

**Kaveh Ahadi and Susanne Stemmer[a)]**

Materials Department, University of California, Santa Barbara, California 93106-5050, USA

[a)] Corresponding author.  Email: stemmer@mrl.ucsb.edu

**Abstract**

We report on a metal-insulator transition (MIT) that is observed in an electron system at the SmTiO$_3$/SrTiO$_3$ interface. This MIT is characterized by an abrupt transition at a critical temperature, below which the resistance changes by more than an order of magnitude. The temperature of the transition systematically depends on the carrier density, which is tuned from ~ $1\times10^{14}$ cm$^{-2}$ to $3\times10^{14}$ cm$^{-2}$ by changing the SmTiO$_3$ thickness. Analysis of the transport properties shows non-Fermi liquid behavior and mass enhancement as the carrier density is lowered. We compare the MIT characteristics with those of known MITs in other materials systems and show that they are distinctly different in several aspects. We tentatively conclude that both long range Coulomb interactions and the fixed charge at the polar interface are likely to play a role in this MIT. The strong dependence on the carrier density makes this MIT of interest for field-tunable devices.

Metal-to-insulator transitions (MITs) that are caused by strong electron correlations are the source of some of the most interesting phenomena in condensed matter physics [1-4]. They can occur either at very high carrier densities, where on-site Coulomb interactions are strong ("strongly correlated systems"), or at ultralow densities, where long-range Coulomb interactions remain unscreened. The corresponding insulating states are the Mott insulator and the Wigner crystal, respectively. Real materials are often significantly removed from either one of these ideal states and contain additional, complex interactions. Furthermore, disorder, which even without correlations can cause localization ("Anderson insulator") [5], can play a central role [6,7]. Transition metal oxides with metallic densities (~one electron/primitive unit cell) have been central to the study of Mott insulators [2], whereas low-dimensional, low-density (sheet densities on order of $10^{11}$ cm$^{-2}$ and below) semiconductors are employed in the search for a Wigner crystal in the solid state [8].

Electron systems at oxide interfaces offer access to a parameter space that is not easily available in the systems mentioned above. Unlike doped or alloyed bulk transition metal oxides, electrostatic doping (for example by a polar discontinuity [9-11]) avoids dopant atoms, a source of disorder. Unlike two-dimensional electron gases in semiconductors, electron masses are high, which promotes localization. Spin-orbit coupling via Rashba fields can also be strong, which may enhance interactions [12]. There are, however, significant challenges. In particular, low carrier mobilities result in high sheet resistances. If these exceed the Mott-Ioffe-Regel limit or quantum resistance ($h/e \sim 25$ k$\Omega$/ ), systems are insulating at most temperatures. This insulating state preempts correlation-induced MITs or other phenomena that may emerge out of an itinerant correlated electron system. MITs near the Mott-Ioffe-Regel limit have been observed in electron systems at LaAlO$_3$/SrTiO$_3$ [13] and SmTiO$_3$/SrTiO$_3$ interfaces [14].

Here, we study SmTiO$_3$/SrTiO$_3$ interfaces that have sheet resistances well below the quantum limit. The polar discontinuity at such interfaces corresponds to a fixed positive charge of 3×10$^{14}$ cm$^{-2}$ at the interface (Fig. 1), which can be compensated by free (mobile) carriers in the SrTiO$_3$ [15]. Mobile carrier densities of 3×10$^{14}$ cm$^{-2}$ per interface are observed for a wide range of sample geometries [11,16]. When SmTiO$_3$ layers are reduced to a few unit cells (u.c.s), the sheet carrier densities are reduced below this value and depend systematically on the SmTiO$_3$ thickness [14]. Given carrier densities that are not close to either one of the two limiting cases discussed above, conventional wisdom would suggest that a correlation driven MIT is unlikely to occur. Contrary to this expectation, we show that an abrupt MIT occurs in this interfacial electron system as the temperature is lowered. A transition temperature that scales systematically with the carrier density and other characteristics point to the importance of electron correlations.

Epitaxial SrTiO$_3$ layers (thicknesses of 20, 60 and 80 nm) were grown using molecular beam epitaxy on (001) (La$_{0.3}$Sr$_{0.7}$)(Al$_{0.65}$Ta$_{0.35}$)O$_3$ (LSAT) single crystals. Following SrTiO$_3$ growth, epitaxial SmTiO$_3$ layers with thicknesses of 3, 5, 7, and 20 u.c.s, respectively, were deposited in-situ (all thicknesses refer to the pseudocubic u.c. parameter of SmTiO$_3$ ~ 3.91 Å). Details of the growth have been reported elsewhere [17,18]. Electron beam evaporation through a shadow mask was used to deposit Au/Ti (400/40 nm) contacts for Hall and sheet resistance ($R_s$) measurements using square Van der Pauw structures. Temperature ($T$) dependent electrical measurements were carried out using a Quantum Design Physical Property Measurement System (PPMS).

Figure 2(a) shows the temperature dependence of $R_s$ of several different heterostructures. All samples, except the structure with 20 u.c.s of SmTiO$_3$, which has nearly the full (3×10$^{14}$ cm$^-$

$^2$) carrier density, exhibit abrupt MITs, with resistance changes exceeding an order of magnitude. [The sample with 20 u.c.s of SmTiO$_3$ only shows a small upturn at low temperatures that could be due to weak localization but is not a sharp MIT as in the other samples]. In most cases, the resistances in the insulating states quickly exceeded the measurement limit and the actual resistance changes are likely much larger. Measurements taken during heating resulted in the same characteristics (no hysteresis was observed), showing reproducible MITs. The MIT temperature ($T_{MIT}$) was sample-dependent and varied from ~40 K (7 u.c.s SmTiO$_3$/60 nm SrTiO$_3$) to near room temperature (260 K for the 3 u.c.s SmTiO$_3$/20 nm SrTiO$_3$ sample). Furthermore, all structures that were still metallic at 110 K showed a hump in $R_s$ at that temperature. This resistance anomaly is more visible in the derivative, d$R_s$/d$T$ vs. $T$, shown in Fig. 2(b) (see arrow). From Fig. 2(b) we see that the anomaly occurs for all samples at ~ 110 K, irrespective of $T_{MIT}$. Thus, the two features are independent phenomena. Bulk SrTiO$_3$ has an antiferrodistortive transition at ~ 110 K and resistance anomalies have been observed to be associated with it [19-21], although their precise origin remains to be determined.

Figure 3 shows $T_{MIT}$ as a function of the sheet carrier density ($n_s$), determined at room temperature from the Hall measurements (shown in Fig. 4). The room temperature $n_s$ values systematically scale with the SmTiO$_3$ thickness [14]. They are 1.4×10$^{14}$, 2×10$^{14}$, 2.1×10$^{14}$ cm$^{-2}$, and 2.7×10$^{14}$ cm$^{-2}$ for the samples with 60 nm SrTiO$_3$ and 3, 5, 7 and 20 u.c.s of SmTiO$_3$, respectively, see Fig. 4(a). Here, $T_{MIT}$ was defined as temperature where d$R_s$/d$T$ ~ 0. Figure 3 shows that $T_{MIT}$ systematically and strongly depends on $n_s$. Even samples with different SrTiO$_3$ thicknesses, which have slightly different mobilities and/or carrier densities than those with 60 nm SrTiO$_3$, follow this general trend. Thus, the primary factor in determining $T_{MIT}$ is the carrier density.

Figure 4 shows the results from Hall measurements. Figure 4(a) shows $(eR_H)^{-1}$, where $R_H$ is the Hall coefficient and $e$ the elementary charge, as a function of temperature. In a simple metal with a single carrier type, $(eR_H)^{-1}$ corresponds to $n_s$. As mentioned above, $n_s$ systematically scales with the SmTiO$_3$ thickness. As can be seen from the large error bars, below $T_{MIT}$, the samples became too resistive to obtain reliable measurements of $R_H$. The slight temperature dependence of $(eR_H)^{-1}$ above the transition is a consequence of the different temperature dependencies of the scattering rates that enter $R_s$ and the Hall angle, $H\cot(\theta_H) = R_s/R_H$ ($H$ is the magnetic field), respectively. For example, $R_s$ is $\sim T^{5/3}$ for the 20 u.c. SmTiO$_3$/60 nm SrTiO$_3$ sample [see dashed line in Fig. 2(b)], whereas $H\cot(\theta_H) \sim T^2$ [see Fig. 4(b)]. Since $R_H$ is the ratio of the two, this causes a temperature dependence of $R_H$ even when there is no actual change in carrier density. The appearance of two different scattering rates, which cannot easily be explained by within Fermi liquid theory, has become known as lifetime separation [22] and is sometimes thought be a signature of proximity to a quantum critical point [23,24], but no definite explanation exists for it yet [25].

The Hall angle, $H\cot(\theta_H)$ is shown in Fig. 4(b). We note that $H\cot(\theta_H) = \mu^{-1}$, where $\mu$ is the Hall mobility. The Hall angle follows a well-defined $T^2$ behavior to room temperature as described by:

$$H\cot\theta_H = H(C + \alpha T^2), \quad (1)$$

where $C$ is the residual and $\alpha$ is the Hall scattering amplitude. $H\alpha$ increases from $1.2\times10^{-6}$ Vs/cm$^2$/K$^2$ for the 20 u.c. SmTiO$_3$/60 nm SrTiO$_3$ sample to $2\times10^{-6}$ Vs/cm$^2$/K$^2$ for the 3 u.c. SmTiO$_3$/60 nm SrTiO$_3$ sample. The 3 u.c. SmTiO$_3$/60 nm SrTiO$_3$ sample displays slight deviations from $T^2$ behavior, in particular, in the proximity of the MIT.

We next discuss the results. First and foremost, we note that the MIT is very different from the MITs that observed near the Mott-Ioffe-Regel limit (or quantum resistance) in a wide range of materials, such as two-dimensional electron gases in semiconductors [26] or oxide thin films [13,27]. In these cases, the MIT occurs near $R_s \sim 25$ kΩ/, when the mean free path length is on order of the lattice spacing [28-30] (of course, the microscopic origins for the high resistance can be quite complex). The typical behavior near such MITs is insulating (or barely metallic) at all *T*, i.e. there is no abrupt transition from metallic to insulating at a critical temperature. In the samples discussed here, only the 3 u.c. SmTiO$_3$/20 nm SrTiO$_3$ sample with $R_s \sim 18$ KΩ/□ falls likely into this category. This sample has a low mobility in addition to the low carrier density, which causes the high $R_s$. It does show, however, an abrupt MIT and was therefore included in Fig. 3. All other samples show an *abrupt* MIT even though $R_s$ is *an order of magnitude below the Mott-Ioffe-Regel limit*. The insulator emerges from a state that is metallic - note the $T^2$ behavior of the Hall angle – and the insulator very quickly becomes highly resistive. Thus, the origin of the MIT observed here must lie in more unconventional physics.

The MIT is reminiscent of that in correlated materials, such as VO$_2$ [31] or the rare earth nickelates [32]. These materials also show an abrupt MIT as the temperature is lowered. However, the MIT in these materials is accompanied by a reduction of the lattice symmetry and they exhibit a pronounced hysteresis indicative of a first order transition. Here we do not observe a hysteresis, suggesting an electronic origin. The strongly insulating state is also striking. The Hall effect can still be measured well below $T_{MIT}$ in charge ordered systems such as the rare earth nickelates (though it may be non-trivial to interpret) [33].

Carrier densities in the samples here are low compared to strongly correlated materials. A significant portion of the carriers spread over the entire SrTiO$_3$ thickness [11,34,35]. This renders the carrier density at a fraction of the 1 electron/u.c. required of a Mott insulator. Even more crucially, $T_{MIT}$ increases as we reduce the carrier density; this is opposite of what would be expected for physics driven by on-site Coulomb interactions, which should become more pronounced as the carrier density is *increased* towards half-filling. Thus, it is unlikely that short-range Coulomb interactions are at the origin of the MIT in this system.

The key observation is that $T_{MIT}$ increases with decreasing $n_s$. Moreover, the scattering amplitude, $\alpha$, also increases with decreasing $n_s$. This indicates an increase in effective mass, $m^*$, since $\alpha T^2 = \frac{m^*}{e} \Gamma(T^2)$, where $\Gamma$ is a temperature-dependent scattering rate and $e$ the elementary charge. A higher effective mass indicates stronger correlations in samples with lower $n_s$. Given the systematic dependence of these two independent quantities on $n_s$, long range Coulomb interactions should be considered. To first order, their importance is determined by the ratio of the interaction energy ($E_{e-e}$) and the kinetic energy (Fermi energy, $E_F$). This ratio often expressed in terms of the Wigner-Seitz radius, $r_s$, which for a two-dimensional electron system is given as:

$$r_s = \frac{E_{e-e}}{E_F} = \frac{e^2 m^*}{\hbar^2 \varepsilon} \frac{1}{\sqrt{\pi n_s}}, \qquad (2)$$

where $\hbar$ the reduced Planck's constant, and $\varepsilon$ the dielectric constant. Using $m^* = 10$, which is not unrealistic for electrons in the $d_{yz,xz}$-derived bands [36], $\varepsilon = 300$ (undoped SrTiO$_3$ at room temperature [37]), and $n_s = 1\times10^{14}$ cm$^{-2}$, we obtain $r_s \sim 5$, which is not high. Typically, electron systems are considered to be strongly correlated for $r_s \sim 10$, with a value of 37 believed to be needed for the Wigner crystal in two dimensions [38]. However, several factors could increase $r_s$ substantially. First, multiple subbands are occupied in this system [34,39-41]. This will increase $r_s$ by factors corresponding to the subband degeneracy. Secondly, a substantial portion of the fixed charge at the interface remains unscreened if the mobile carrier density is less than $3\times10^{14}$ cm$^{-2}$. We believe the fixed charge at the polar interface is a key ingredient in the microscopic origins of this MIT, as no abrupt MITs with change in temperature are observed either in bulk doped SrTiO$_3$ [42] or in other interfacial systems involving SrTiO$_3$, such as LaAlO$_3$/SrTiO$_3$, at comparable (or even lower) carrier densities. One consequence of the polar fixed charge are very large, asymmetric electric fields, and thus Rashba-type of spin-orbit coupling, which is believed to aid in the localization [12]. Finally, we note that the samples are certainly not free of disorder, which may promote unconventional metallic and insulating states even at small $r_s$ [7].

In conclusion, the observed mass enhancement, non-Fermi liquid behavior, MITs at resistances much below the quantum resistance, and, most of all, the systematic scaling of $T_{MIT}$ with the carrier density all point to an unconventional insulator in which long-range electron correlations play a key role. Future studies, though experimentally challenging, should be directed at determining the nature of insulating state, in particular with regards to magnetic order, or absence thereof, and the spatial distribution of carriers. The results highlight the potential for new discoveries in MITs in materials systems that are substantially different from the well-studied strongly correlated Mott insulators and dilute semiconductor systems. They also have implications for practical applications. The fact that $T_{MIT}$ is highly sensitive to the carrier density makes this system of substantial interest for electric field gating of MITs. The relatively low (in comparison to strongly correlated materials) carrier densities promise an MIT that can be gated with conventional solid-state gate dielectrics, which has been a long-sought-after goal [43].

**Acknowledgements**

The authors thank Leon Balents and Jim Allen for discussions. This work was supported in part by FAME, one of six centers of STARnet, a Semiconductor Research Corporation program sponsored by MARCO and DARPA. The MRSEC Program of the National Science Foundation (Award No. DMR 1121053) supported some of the facilities that were used in this study.

**Figure Captions**

**Figure 1:** Schematic of the SmTiO$_3$/SrTiO$_3$ interface, showing the formal charges carried by the layers. The polar discontinuity corresponds to a fixed charge at the interface, which can be compensated by ½ mobile electrons/interface unit cell, or ~ 3×10$^{14}$ cm$^{-2}$. The SmTiO$_3$ surface has a similar polar problem and various ways of potentially resolving it (adsorbates, nonstoichiometry,…) [15]. For thick SmTiO$_3$, interfaces have the full 3×10$^{14}$ cm$^{-2}$ mobile charge density, but for thin SmTiO$_3$, the carrier density is reduced. It is likely that this is due to the proximity to the polar surface. A significant fraction of the mobile electrons, specifically those that are located in $d_{xz,yz}$ orbitals, spread out far into the SrTiO$_3$ [34,35]. The degree to which the electron gas spreads is determined by a number of boundary conditions, including the carrier density and field-dependent permittivity of SrTiO$_3$. The SrTiO$_3$ thickness thus affects the spatial distribution.

**Figure 2:** (a) $R_s$ as a function of temperature for heterostructures with various SmTiO$_3$ (3, 5, 7, and 20 u.c.s) and SrTiO$_3$ (20, 60, and 80 nm) thicknesses, respectively, as indicated in the legend. (b) Calculated d$R_s$/d$T$ as a function of temperature for the same structures. The arrow indicates the resistance anomaly at 110 K, which is observed for all samples. The dashed line is a fit to a power law, $T^n$, behavior of the 20 u.c. SmTiO$_3$ sample, which determined $n$ ~ 5/3. The limited temperature range above $T_{MIT}$ for the other samples precluded similar fits.

**Figure 3:** $T_{MIT}$ as a function of $n_s$ for the samples shown in Fig. 2. The dotted line is a guide to the eye. The labels indicate the samples, with the first number referring to the SmTiO$_3$ thickness in number of unit cells and the second number to the SrTiO$_3$ thickness in nm. We note that for the sample with 20 u.c.s of SmTiO$_3$, which does not show an abrupt MIT, the temperature of the weak upturn in the resistance is shown instead.

**Figure 4:** (a) $(eR_H)^{-1}$ as a function of temperature. Data below 50 K is not shown as the resistances became too high for reliable Hall measurements for all samples except the one with 20 u.c. of SmTiO$_3$ (note the very large error bars at low temperatures). (b) $H\cot(\theta_H)$ as a function of $T^2$. The lines are a fit to $T^2$ behavior that is used to obtain $H\alpha$, according to Eq. (1).

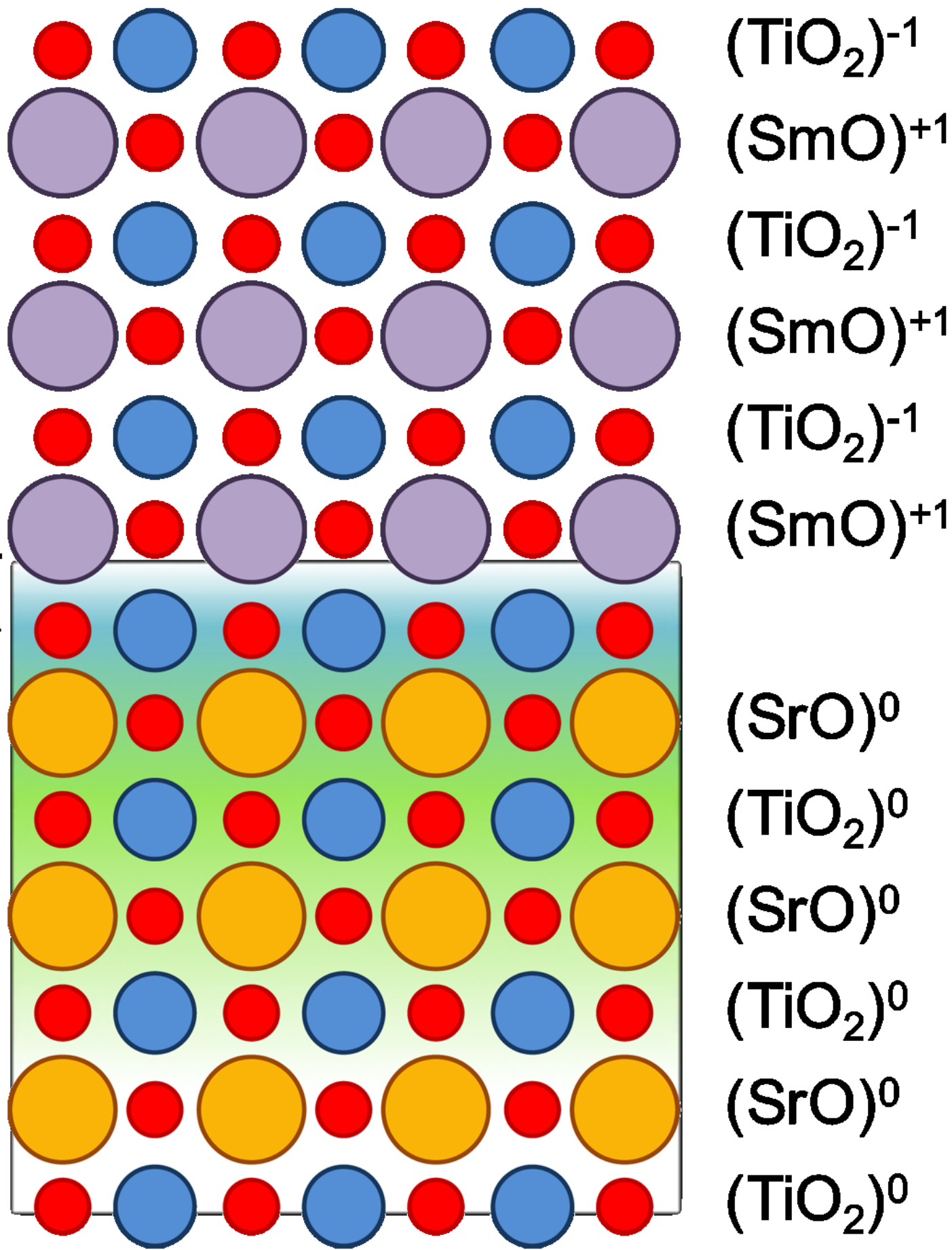

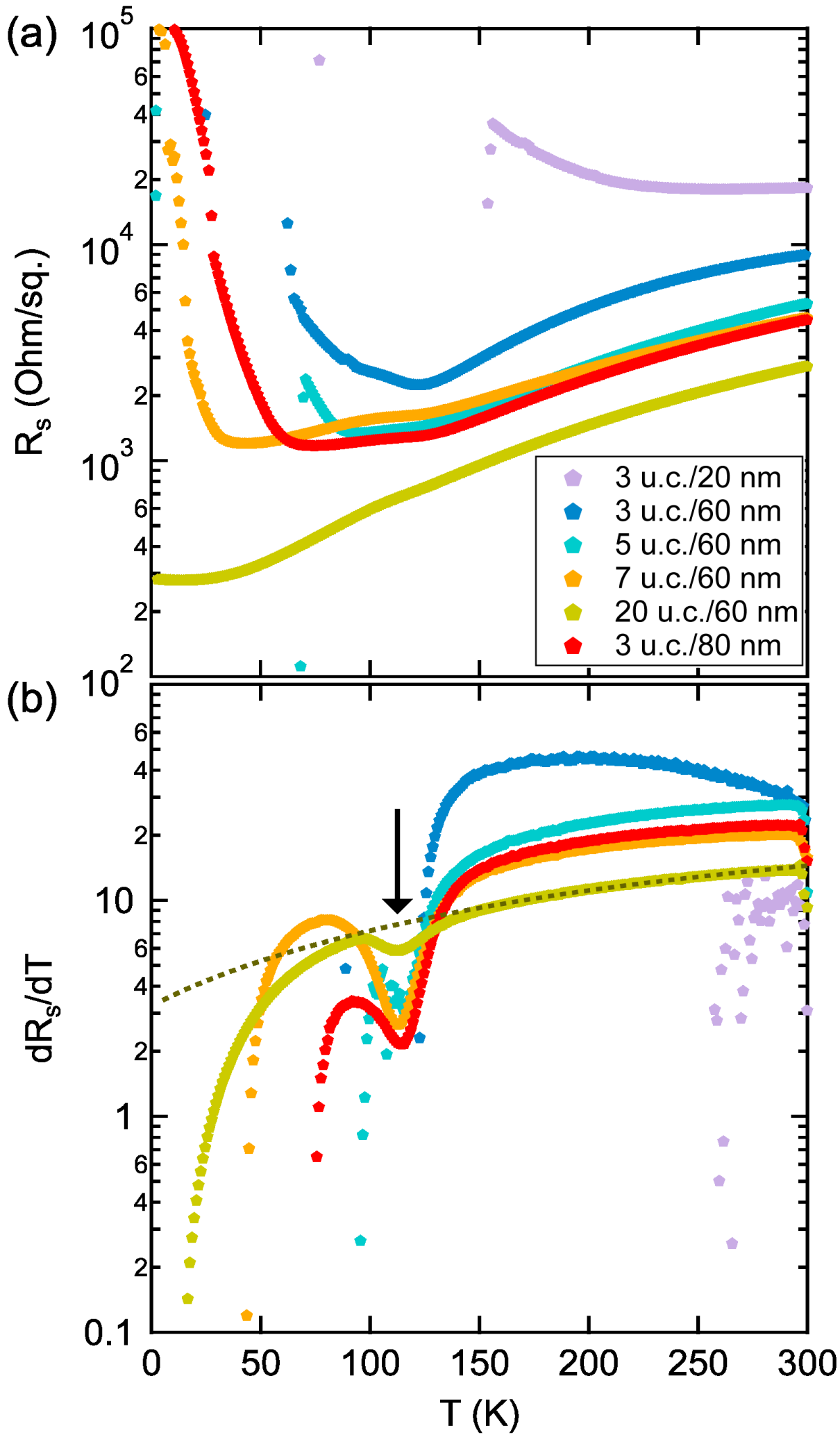

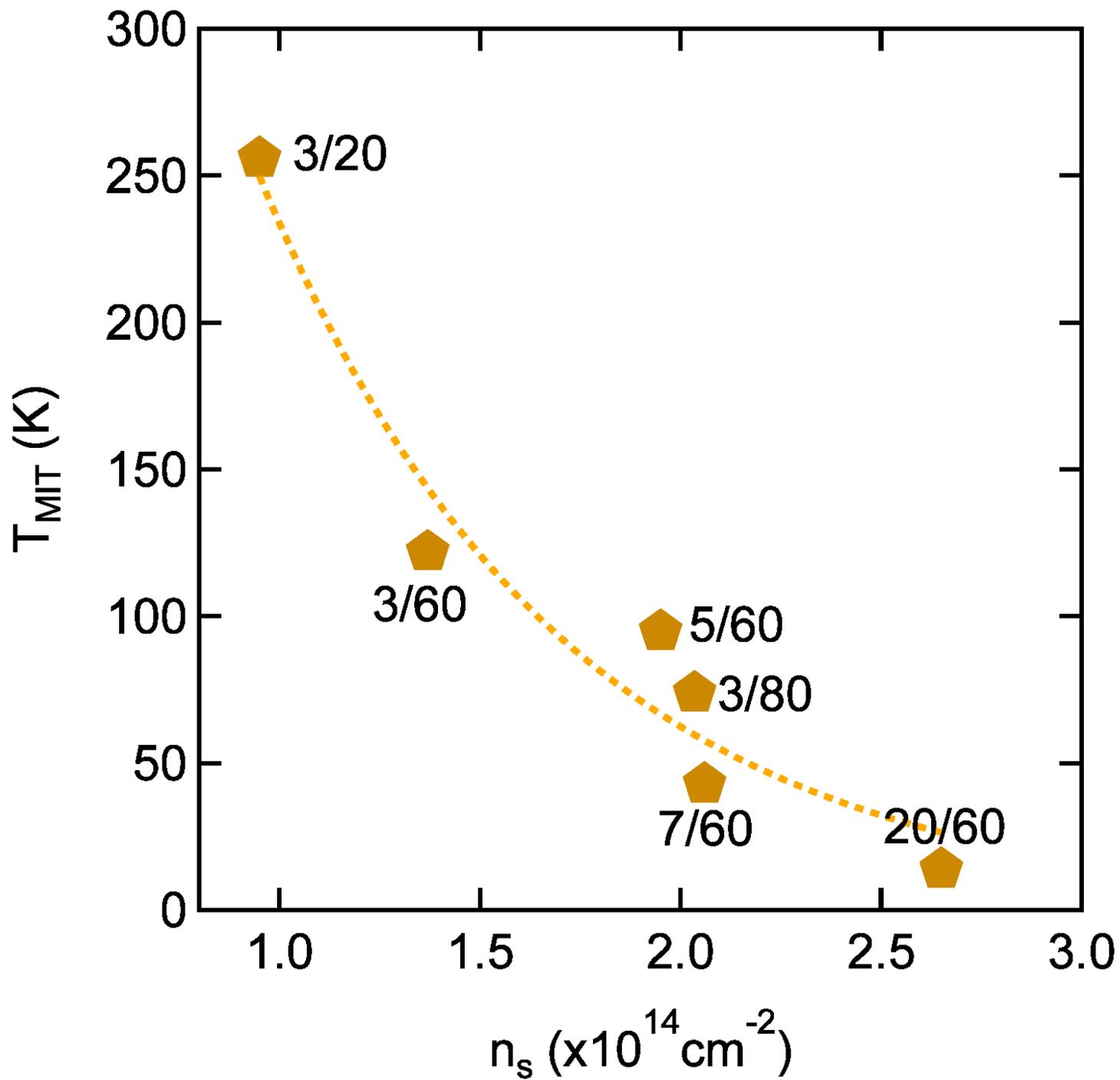

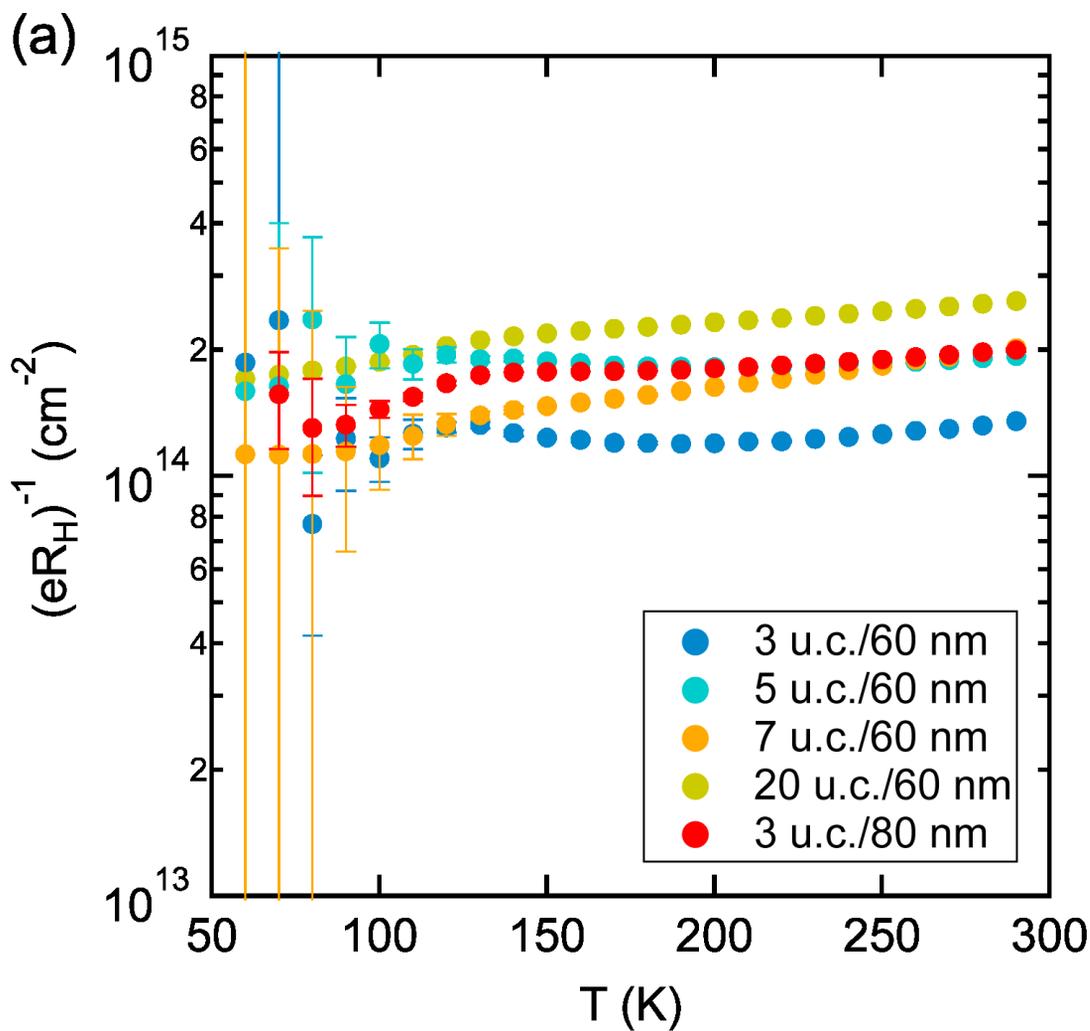

(a)

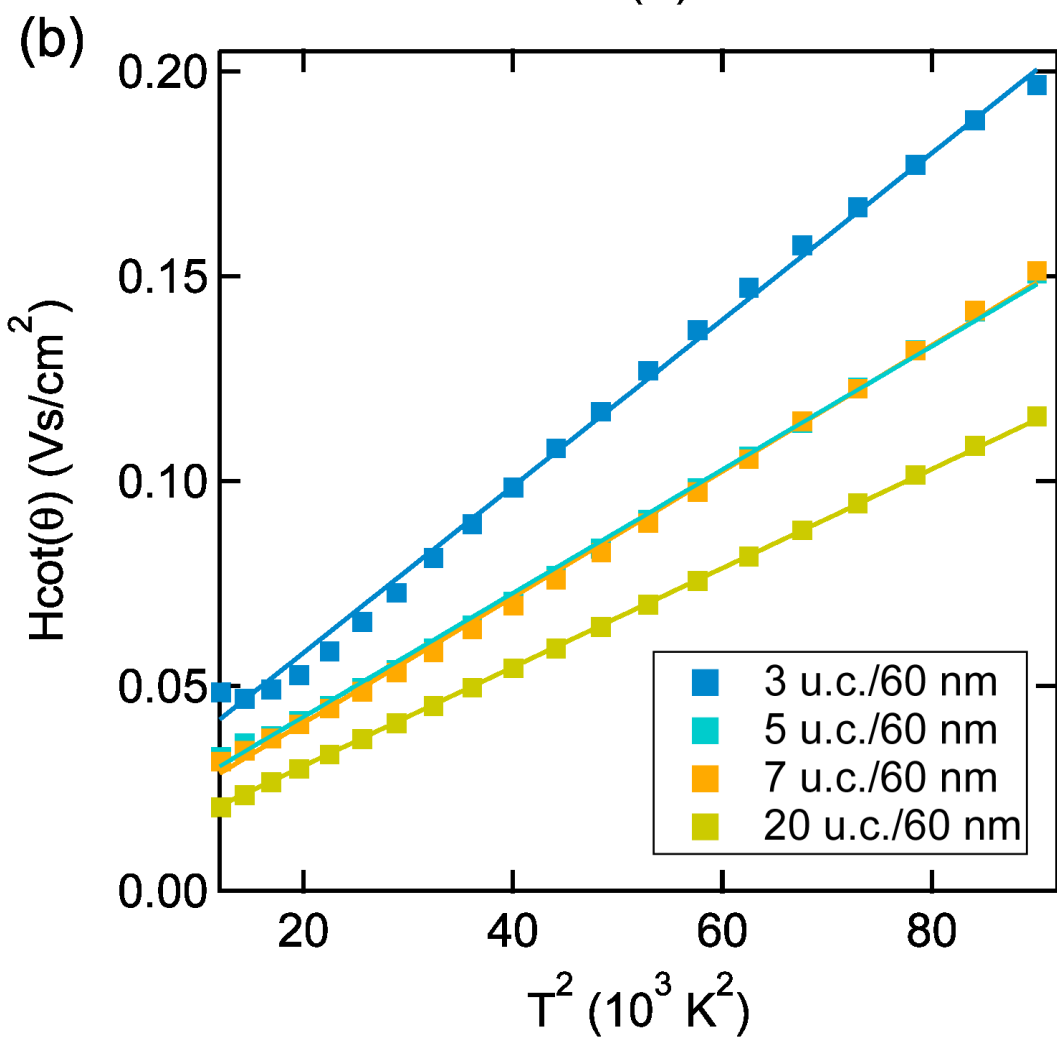

(b)